\documentclass[conference]{IEEEtran}
\ifCLASSOPTIONcompsoc
  \usepackage[nocompress]{cite}
\else
  \usepackage{cite}
\fi
\ifCLASSINFOpdf
\else
\fi
\usepackage{amsmath}
\usepackage{algorithm2e}
\usepackage{algorithmic}

\usepackage{array}

\ifCLASSOPTIONcompsoc
  \usepackage[caption=false,font=footnotesize,labelfont=sf,textfont=sf]{subfig}
\else
  \usepackage[caption=false,font=footnotesize]{subfig}
\fi
\usepackage{url}

\usepackage{mathtools}
\DeclarePairedDelimiter\ceil{\lceil}{\rceil}

\usepackage{textcomp}
\usepackage{xcolor}

\usepackage{booktabs}
\usepackage{comment}

\usepackage{titlesec}
\titlespacing*{\section}{0pt}{*1}{*1}
\titlespacing{\subsection}{0pt}{*1}{*1}

\newcommand{\ours}{\textsf{AdaFed}}

\begin{document}
%
\title{Adaptive Aggregation For Federated Learning}


\author{\IEEEauthorblockN{K. R. Jayaram},
\IEEEauthorblockA{IBM Research, USA} \and
\IEEEauthorblockN{Vinod Muthusamy},
\IEEEauthorblockA{IBM Research, USA} \and
\IEEEauthorblockN{Gegi Thomas},
\IEEEauthorblockA{IBM Research, USA} \and
\IEEEauthorblockN{Ashish Verma},
\IEEEauthorblockA{IBM Research, USA} \and
\IEEEauthorblockN{Mark Purcell},
\IEEEauthorblockA{IBM Research, Ireland}
}


%


\maketitle

\begin{abstract}

In this paper, we present a new scalable and adaptive architecture for FL aggregation.
First, we demonstrate how traditional tree overlay based aggregation techniques 
(from P2P, publish-subscribe and stream processing research) can help FL aggregation scale,
but are ineffective from a resource utilization and cost standpoint. Next, we present the 
design and implementation of \ours, which uses serverless/cloud functions to adaptively 
scale aggregation in a resource efficient and fault tolerant manner. 
We describe how \ours\ enables FL aggregation to be dynamically deployed only when necessary,
elastically scaled to handle participant joins/leaves and is fault tolerant with minimal effort
required on the (aggregation) programmer side. We also demonstrate that our prototype based on Ray~\cite{ray}
scales to thousands of participants, and is able to achieve a $>90\%$ reduction in resource requirements and cost, 
with minimal impact on aggregation latency.


\end{abstract}


\section{Introduction}

Federated Learning (FL)~\cite{kairouz2019advances, fieldguide}  is a mechanism in which
multiple parties collaborate to build and train a joint machine learning model typically 
under the coordination/supervision of a central server or service provider (definition
by Kairouz et. al.~\cite{kairouz2019advances, fieldguide}). This central server is also
called an \emph{aggregator}. FL is private by design, because parties retain their 
data within their private devices/servers; never sharing said data with either
the aggregator or other parties. An FL job involves parties performing local training on their data, 
sharing the weights/gradients of their model (also called a \emph{model update}) with the aggregator, 
which aggregates the model updates of all parties using a fusion algorithm. 
The use of \emph{centralized aggregation} is common in FL because of the ease in which
various machine learning models (neural networks, decision trees, etc.) and 
optimization algorithms can be supported.

FL is \emph{typically} deployed in two scenarios: \emph{cross-device} and \emph{cross-silo}.
In the cross-silo scenario, the number of parties is small, but each party has extensive 
compute capabilities (with stable access to electric power and/or equipped with hardware accelerators) 
and large amounts of data. The parties have reliable participation throughout the entire federated 
learning training life-cycle, but are more susceptible to sensitive data leakage. Examples include 
multiple hospitals collaborating to train a tumor/COVID detection model on radiographs~\cite{nvidia-covid}, multiple banks 
collaborating to train a credit card fraud detection model, etc.
The cross-device scenario involves a large number of parties ($>100$), but each party has a small 
number of data items, constrained compute capability, and limited energy reserve (e.g., mobile phones or IoT devices). 
They are highly unreliable/asynchronous and are expected to drop and join frequently. Examples include a large 
organization learning from data stored on employees' devices and a device manufacturer training a model 
from private data located on millions of its devices (e.g., Google Gboard~\cite{bonawitz2019towards}). 

Increasing adoption of FL has, in turn, increased the need for 
FL-as-a-service offerings by public cloud providers, which serve as a nexus 
for parties in an FL job and aggregate/fuse model updates. 
Such FL aggregation services have to effectively support multiple concurrent 
FL jobs, with each job having tens to thousands of heterogeneous participants (mobile phones,
tablets, sensors, servers) from different organizations and administrative domains.
Our experience, in building and operating the IBM Federated Learning (IBM FL)~\cite{ibmflpublic, ibmfl} 
 service on our public and private clouds has led us to believe that existing FL aggregation
 methods have performance, scalability and resource efficiency challenges, primarily 
 due to the use of centralized aggregation.

\noindent{\bf Performance:} Aggregators should not become a bottleneck or a single point 
 	of failure in FL jobs. They should be able to store incoming model updates without loss,
 	and have low latency -- the time between the arrival of the last expected model update
 	and the completion of aggregation. In the case of a cloud hosted FL aggregation service,
 	said guarantees must hold across all running FL jobs. Most existing FL platforms 
(IBM FL~\cite{ibmfl}, Webank FATE~\cite{fate}, NVIDIA NVFLARE~\cite{nvflare}) are based on a client-server model with 
a single aggregator per FL job deployed (as a virtual machine or container) 
in datacenters waiting for model updates. Such platforms are able to easily support multiple concurrent
FL jobs, but performance drops as the number of parties increases, especially in cross-device settings. 
This is because aggregation 
throughput is limited by the computational capacity of the largest VM or container 
(memory and compute, and to a lesser extent, network bandwidth).

\noindent{\bf Scalability:} is considered in terms of the number of parties, size of model
updates, frequency of updates and (for an FL service) number of concurrent FL jobs. FL platforms
using a single aggregator per job only support vertical scalability; non-trivial design
using data parallelism and connecting multiple aggrgeators 
is necessary for horizontal scalability, especially in cross-device settings. FL jobs involve several rounds,
and take an extended period of time, especially with intermittently available parties. Party joins and
dropouts are common; so aggregation infrastructure must scale horizontally to support this.

\noindent{\bf Resource Efficiency/Cost:} While operating IBM FL and from publicly available FL benchmarks
like LEAF~\cite{leaf-benchmark} and Tensorflow Federated~\cite{tff-benchmark}, we have observed that
 training at the party
takes much longer compared to model update fusion/aggregation, resulting in under-utilization and 
wastage of computing resources dedicated to aggregation. This is a significant problem even 
in cross-silo settings -- active participation is not guaranteed even in cross-silo settings 
due to competition from other higher priority workloads and variations in data availability.
It is further compounded in ``cross-device'' deployments, where  parties are highly \emph{intermittent} 
and do not have dedicated resources for training. 
In these scenarios, the 
    aggregator expects to hear from the parties \emph{eventually} (typically over a several hours or maybe once
    a day). Large-scale FL jobs almost always involve intermittent parties -- as the number
    of parties increases, it is extremely hard to expect that all of them participate at the same pace.
This results in aggregators having to wait for long periods of time for parties to finish local 
training and send model updates.

\noindent {\bf Contributions:} The core technical contribution of this paper is the design, implementation and evaluation of a 
flexible parameter aggregation mechanism for FL -- \ours, which has the following novel features:

\begin{itemize}
   \item \ours\ reduces state in aggregators and treats aggregators as serverless functions. In many existing FL jobs,
    every aggregator instance typically 
     acts on a sequence of inputs and produces a single output. State, if present, is not local to the aggregator 
     instance and may be shared by all aggregators. Such state is best left in an external store, and consequently 
     aggregators can be completely stateless and hence, serverless. \ours\ is therefore scalable both with respect to 
     participants -- effective for cross-silo and cross-device deployments, 
     and with respect to geography -- single/hybrid cloud or multicloud.
    
     \item \ours\ leverages serverless technologies to deploy and tear down aggregator instances dynamically 
     in response to participant model updates, thereby supporting both intermittent and active participants
     effectively. There is no reason to keep aggregators deployed all the time and simply ``awaiting input''.
    
    \item \ours\ is efficient, both in terms of resource utilization with support for automatic elastic scaling, and in terms of aggregation latency.
    \item \ours\ is reasonably expressive for programmers to easily implement scalable aggregation algorithms.  
    \ours\ is implemented using the popular Ray~\cite{ray} distributed computing platform, and can run arbitrary Python code
    in aggregation functions, and use GPU accelerators if necessary.

    \item Increased FL job reliability and fault tolerance by reducing state in aggregators, eliminating persistent 
    network connections between aggregators, and through dynamic load balancing of participants.

    \item \ours\ supports widely used FL privacy preserving and security mechanisms

\end{itemize}


\section{Background : FL Aggregation}~\label{sec:background}

\begin{algorithm}
\small
       \caption{Generalized FedAVG~\cite{fieldguide}}~\label{alg:fedavg}
        \mbox{{\bf Aggregator Side}} \\ 
        \mbox{  } \\
        $\mbox{Initial model } m^{1}$ \\
        \For(){$r ~\in~ \{1,2,\ldots,R\}$}{
            Sample a subset $\mathcal{S}^{(r)}$ of participants \\
            \textsc{send} $m^{(r)}$ \mbox{to each} $i \in  \mathcal{S}^{(r)}$ \\
            \textsc{recv} \mbox{model update} $\triangle_{i}^{(r)}$ \mbox{from each} $i \in  \mathcal{S}^{(r)}$ \\
            \mbox{Aggregate } $\triangle^{(r)} \gets \frac{1}{N} {\sum_{i \in  \mathcal{S}^{(r)}} n_i \triangle_{i}^{(r)}}$ \\
            $m^{(r+1)} \gets \textsc{optimizer} (m^{r}, -\triangle^{(r)}, \eta^{(r)})$
        }
        \mbox{  } \\
        \mbox{{\bf Participant Side}} \\ 
        \textsc{recv} ($m^{(r)}$) from aggregator \\
        Local model $x^{(r,1)} \gets m^{(r)}$\\
        \For(){$k \in \{1,2,\ldots,\tau \} $ } {
          Compute local stochastic gradient $g_i(x^{(r,k)})$\\
          $x^{(r,k+1)} \gets \textsc{optimizer}(x^{{r,k}}, -g_i(x^{(r,k)}), \eta^{(r)})$\\
         }
         Compute local model update $\triangle^{(r,l)} \gets x^{(r,\tau)} - x^{(r,1)}$\\
         \textsc{send} $\triangle^{(r,l)}$ to aggregator \\
         \mbox{  } \\
        
        \end{algorithm}

An aggregator typically coordinates the entire FL job.
The parties, aided by the aggregator, agree on the model architecture (ResNet, EfficientNet, etc), 
optimizer to use (SGD, Adam, AdaGrad, etc.) and hyperparameters to be 
used for the FL job (batch size, learning rate, aggregation frequency etc.). The aggregator is responsible for 
durably storing the global model and keeping track of the FL job.
We illustrate FL using the most common algorithm used for neural networks and 
gradient descent based machine learning models -- FedAvg~\cite{fieldguide}.
For FedAvg (Algorithm \ref{alg:fedavg}), the aggregator selects a random subset $\mathcal{S}^{(r)} \subset \mathcal{S}$ of parties 
for every round $r$. The aggregator initializes the global model $m^{1}$ using the same process as if the
job is centralized (i.e, either randomly or from existing pre-trained models). At each round, the 
aggregator transmits the global model $m^{(r)}$ to $\mathcal{S}^{(r)}$. Once a party receives $m^{(r)}$, it uses
$m^{(r)}$ to make $\tau$ training passes on its local dataset. $\tau$ is the aggregation frequency.
It then computes the local gradient update
after $\tau$ passes, $\triangle^{(r,l)}$, and transmits the same to the aggregator. The aggregator in FedAvg
then computes the weighted average of all gradient updates -- 
$\frac{1}{N} \sum_{i \in  \mathcal{S}^{(r)}} n_i \triangle_{i}^{(r)}$ to 
compute the global gradient update $\triangle^{(r)}$ and update the global model (for the next round)
$m^{(r+1)}$. This process proceeds for a set number $R$ of rounds or until the aggregator 
has determined that the model has converged. The term $n_i$ in the weighted average is the number of training 
samples at party $i$ and $N$ is the total number of training samples involved in the round, i.e., $N = \sum_{i \in  \mathcal{S}^{(r)}} n_i$.

\noindent{\bf Associativity of Aggregation:}  
Since the number of participants typically varies between FL jobs, 
and within a job (over time) as participants join and leave, horizontal scalability of FL
aggregation software is vital. \emph{Horizontally scalable} aggregation is only feasible 
if the aggregation operation is associative -- assuming $\oplus$ denotes the aggregation of model updates (e.g., gradients)
$U_i$, $\oplus$ is associative if $U_1 \oplus U_2 \oplus U_3 \oplus U_4 \equiv (U_1 \oplus U_2) \oplus (U_3 \oplus U_4)$.
Associativity is the property that enables us to exploit data parallelism to 
partition participants among aggregator instances, with 
each instance responsible for handling updates from a subset of participants.
The outputs of these instances must be further aggregated.
In the case of FedAvg, $\sum_{i \in  \mathcal{S}^{(r)}} n_i \triangle_{i}^{(r)}$ is associative because addition is associative,
and the most computationally intensive because each $\triangle_{i}^{(r)}$ involves millions of floating point numbers.
A common design pattern in parallel computing~\cite{grama} is to use tree-based or hierarchical aggregation 
in such scenarios, with a tree topology connecting the aggregator instances. The output of each aggregator
goes to its parent for further aggregation.

\section{\ours\ : Design and Implementation}

\begin{figure}[htb]
    \centering
    \includegraphics[width=0.8\columnwidth]{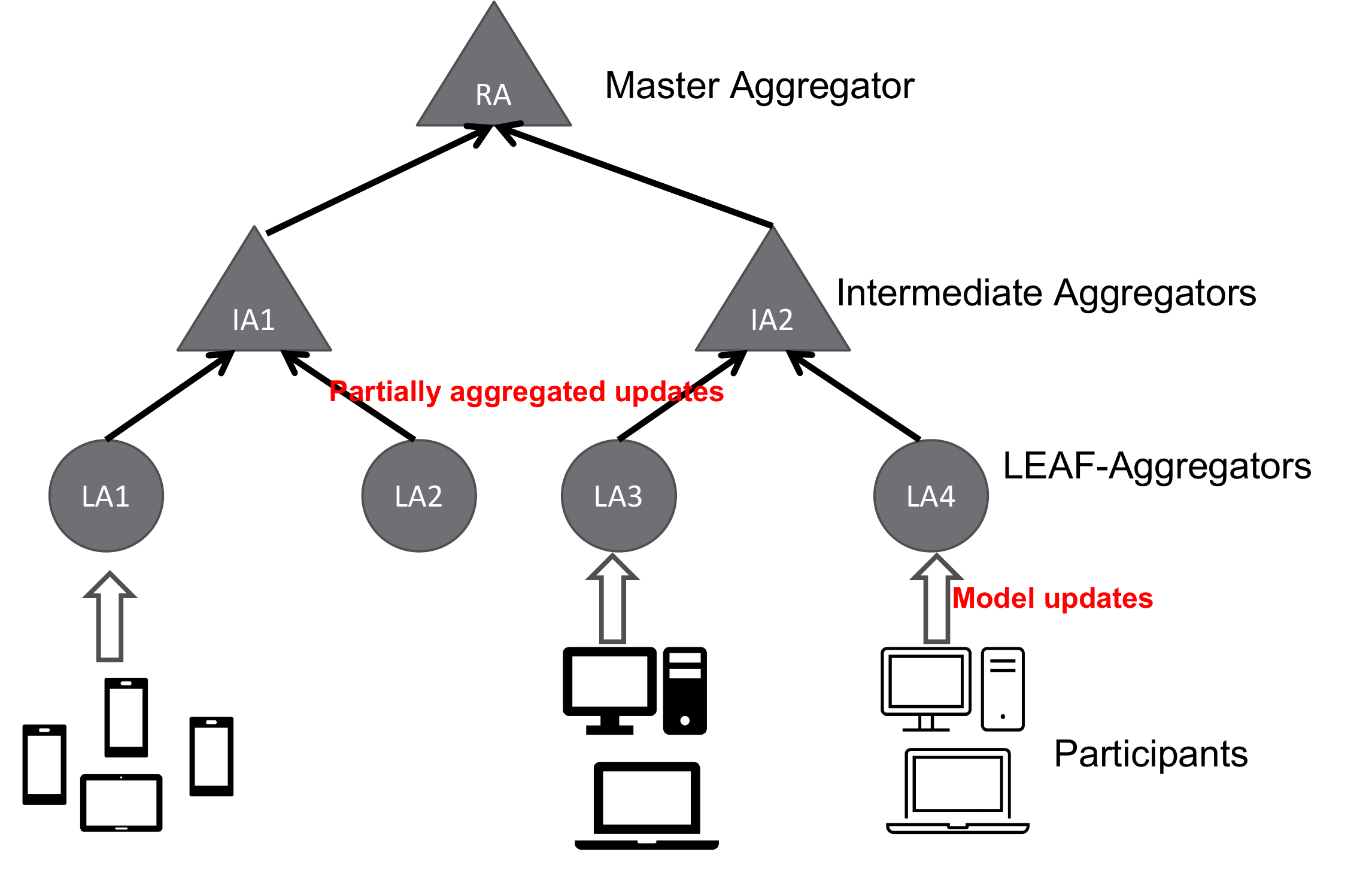}
    \caption{Hierarchical/Tree-based Aggregation}
    \label{fig:hierarchical}
\end{figure}

\ours, as its name suggests, adapts to the mechanics of a specific FL job.
When a job's aggregation function is associative, as it is in most FL jobs,
 \ours\ leverages data parallelism
to spawn several aggregation ``entities/instances'' per FL job and arranges them 
in a tree based (hierarchical) overlay. Tree-based overlays are a common distributed computing
pattern in publish-subscribe~\cite{pubsub} and stream processing~\cite{streamoverlays}.
This enables aggregation to scale to support thousands of parties. 
However, using ``statically deployed'' (always on) overlays, while advantageous in 
high throughput stream processing, is not suitable for FL. 

Consequently, \ours\ has a programming model whose
goal is to reduce state in aggregators and to decouple aggregator instances.
This enables said instances to execute as serverless functions, which are spawned only
when model updates arrive, and are torn down when parties are busy training (no updates available
to aggregate). An aggregation function instance can be triggered once a specific number of model updates
are available; or multiple instances can be triggered once the expected number of model updates for the 
current FL round are available. Once a model aggregation round is complete and the fused model 
is sent back to the parties, all aggregator functions exit until the next round, 
thereby releasing resources.

\subsection{Associativity $\rightarrow$ Tree-based Aggregation}

Associativity enables us to partition parties among aggregator instances, with 
each instance responsible for handling updates from a subset of parties.
The outputs of these instances must be further aggregated.
A tree topology connects the aggregator instances. The output of each aggregator
goes to its parent for further aggregation. We have determined that it is possible to split any 
associative FL aggregation operation into leaf and intermediate aggregators as illustrated by
Figure~\ref{fig:hierarchical}. A leaf aggregator implements logic to fuse raw model weight updates $U_i$ from a group of 
$k$ parties to generate a partially aggregated model update $U_k$. For example, in the case of 
FedAvg~\cite{mcmahan2017communication, mcmahan2017learning} this function would take 
$k_i$ gradient update vectors and return the weighted sum $S_i = \sum_{1,\ldots,k_i} n_i \triangle_{i}^{(r)}$ of these vectors,
along with the number of data items processed so far $\sum_{1,\ldots,k_i} n_i$. 
An intermediate aggregator implements logic to further aggregate partially 
    aggregated model updates ($U_k$), in stages, to produce the final aggregated model 
    update ($U_F$). In the case of FedAvg, this function would aggregate (add up) multiple
    $(S_i)$. If all expected model updates have arrived from $\mathcal{S}^{(r)}$ parties, the 
    intermediate aggregator would have thus calculated $\sum_{1,\ldots,|\mathcal{S}^{(r)}|} n_i \triangle_{i}^{(r)}$
    and $N=\sum_{1,\ldots,|\mathcal{S}^{(r)}|} n_i$, from which the aggregated gradient update $\triangle^{(r)}$ 
    is calculated per Algorithm~\ref{alg:fedavg} at the root/master aggregator (Figure~\ref{fig:hierarchical}).

Establishing a tree-based aggregation topology as in Figure~\ref{fig:hierarchical}
starts by identifying the 
number of parties that can be comfortably handled by an aggregator instance.
This is dependent on (i) size/hardware capability (CPU/RAM/GPU) of the instance (server or VM or container) and
its network bandwidth, and (ii) the size of the
model, which directly determines the size of the model update and the memory/compute 
capabilities needed for aggregation. Assuming that each instance can handle $k$ participants,
a complete and balanced $k$-ary tree can be used. $\ceil{\frac{n}{k}}$ leaf aggregators are 
needed to handle $n$ participants; the tree will have $O(\ceil{\frac{n}{k}})$ nodes.

While a tree-based FL aggregation overlay is conceptually simple, it does involve significant
implementation and deployment effort for fault tolerant aggregation. 
Typically, aggregator nodes are instantiated 
using virtual machines (VMs) or containers (e.g., Docker) and managed using
a cluster management system like Kubernetes. These instances are then arranged in the 
form of a tree, i.e., each instance is provided with the IP address/URL of its parent,
expected number of child aggregators, credentials to authenticate itself to said parent and 
send aggregated model updates. Failure detection and recovery 
is typically done using heartbeats and timeouts, between each instance, its parents and children.
Once faults happen, the aggregation service provider should typically take responsibility for
recovering the instance, and communicating information about the recovered instance to 
its children for further communications. Things become complicated when an instance fails
at the same time as one of its parent or child instances. Another issue, common in distributed software systems, that arises in this scenario is 
network partitions. In summary, to implement hierarchical 
aggregation the traditional way~\cite{grama}, any aggregation service 
has to maintain dedicated microservices to deploy, monitor and heal these aggregation overlays.


\subsection{``Idle Waiting'' in Static Tree Aggregation}

Even if some technologies like Kubernetes pods and service abstractions are able to simplify
a few of these steps, a more serious problem with tree-based aggregation overlays is that 
aggregator instances are ``always on'' waiting for updates, and this is extremely wasteful in terms
of resource utilization and monetary cost.  To handle FL jobs across thousands of parties, 
aggregation services including \ours\ 
must support intermittent parties effectively. Given that, for every round, parties
may send model updates over an extended time period (hours), aggregators spend the bulk
of their time waitin. Idle waiting wastes resources and increases aggregation cost.
A tree-based aggregation overlay compounds resource wastage and cost.

Re-configuring tree-based aggregation overlays is also difficult. This is needed, for example,
when midway through a job, a hundred (or a thousand) participants decide to join. 
Supporting them would require reconfiguration at multiple levels of the aggregation overlay.
Reconfigurations are also necessary to scale down the overlay when participants leave. 
Thus, elasticity of aggregation is hard to achieve in the static tree setting.

\begin{figure}[htb]
    \centering
    \includegraphics[width=0.75\columnwidth]{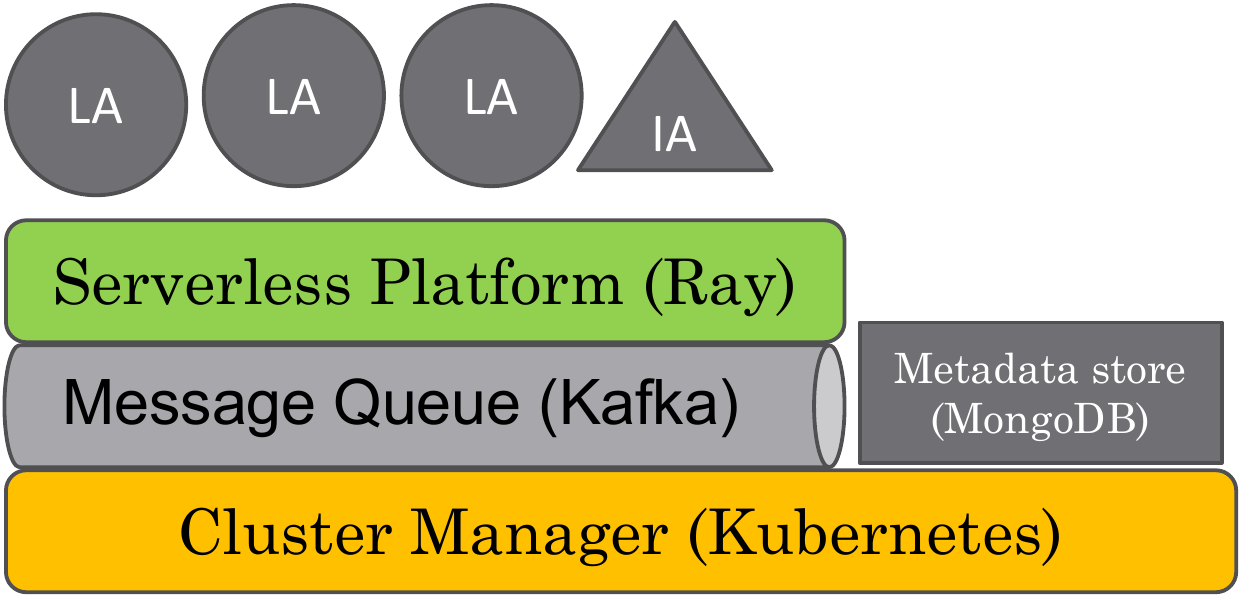}
    \caption{\ours\ System Architecture. Aggregators are executed as serverless functions.}
    \label{fig:lambdafl-arch}
\end{figure}

\begin{figure}[htb]
    \centering
    \includegraphics[width=\columnwidth]{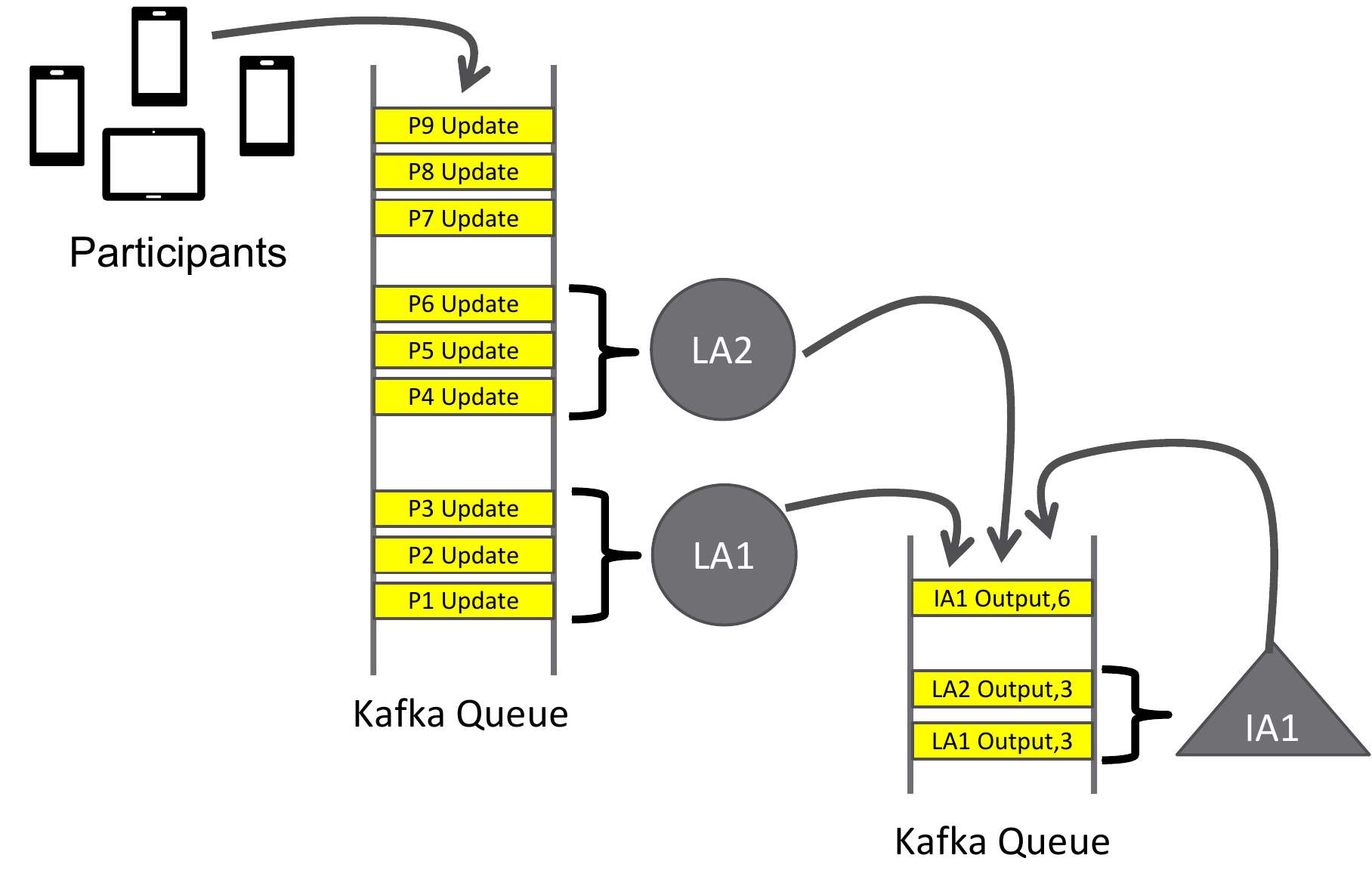}
    \caption{\ours\ -- Illustration of stepwise serverless aggregation}
    \label{fig:qagg}
\end{figure}

\subsection{Using Serverless Functions}

\ours\ takes associativity one step further. \ours\ mitigates issues with aggregation overlays by avoiding the construction of \emph{actual/physical}
tree topology. Instead, \ours\ uses serverless functions chained together with message queues
to realize a \emph{logical} tree topology. \ours\ executes both leaf and intermediate aggregation
operations as serverless/cloud functions. These functions are executed in containers on a 
cluster managed by Kubernetes, which multiplexes 
multiple workloads and enables the cluster to be shared by multiple FL jobs and/or other workloads.
Also, since there is no static topology, more (or less) aggregator functions can be spawned depending
on the number of parties (model updates), thereby handling party joins/leaves effectively.
The challenge in executing aggregation as serverless functions, which are ephemeral and have no 
stable storage, is to manage state -- that of each aggregation entity, intermediate aggregation 
outputs, inter-aggregator communications and party-aggregator communications.
We also note that splitting aggregation into leaf and intermediate 
functions makes the logic simpler. It is also possible to have a single serverless function that can operate
on both raw updates and partially fused updates; doing that will increase the complexity of the function.

\subsection{Party-Aggregator Communication}

This is done using a distributed message queue (Kafka). Kafka is a topic-based
    message queue offering standard publish/subscribe semantics. That is, each queue has a ``name'' (i.e., pertains to a ``topic''), and
    multiple distributed entities can write to (publish) and read from (subscribe to) it. Kafka enables us
    to set a replication level per queue, which ensures durability of messages between the aggregator instances
    and parties. For each FL job (with an identifier \textsf{JobID}, two queues are created at deployment time 
    -- \textsf{JobID-Agg} and \textsf{JobID-Parties}.
    Only aggregator instances (serverless functions) can publish to \textsf{JobID-Agg} and all parties subscribe to it. Any party can publish to
    \textsf{JobID-Parties} but only the aggregator instances can both publish to and read from it. This ensures that model updates
    sent to \textsf{JobID-Parties} are private and do not leak to other parties. When the job starts, the aggregator publishes the 
    initial model on \textsf{JobID-Agg}; parties can then download the model and start training. At the end of each 
    job round, parties publish their model updates to \textsf{JobID-Parties}. \emph{Inter-Aggregator Communication,} 
    is also handled using Kafka. Partially fused model updates are published
    by aggregation functions into Kafka, and can trigger further function invocations.
    
\subsection{Aggregation Trigger}

For serverless functions to execute, they must be triggered by some event. 
    \ours\ provides several flexible and configurable triggers. The simplest ones trigger an aggregation function for
    every $k$ updates published to \textsf{JobID-Parties}, or every $t$ seconds. For FL jobs that use a parameter server strategy
     for model updates,
    it is possible in \ours\ to implement the update logic as a serverless function and trigger it every time 
    an update is published by a party. Other custom triggers involve the periodic execution of any
    valid Python code (also as a serverless function) which triggers aggregation.
    Custom triggers are vital to handling FL jobs involving intermittent parties. As an illustration, consider
    an FL job where each round is successful if 50\% of parties send model updates within 10 minutes. The aggregation trigger here could be a serverless function, invoked every minute, to
    count the number of parties that have responded and perform partial aggregation through leaf
    aggregators; aggregation is complete when at least 50\% of the parties have responded. Another
    FL job may require that aggregation waits for at least 10 minutes and considers the round successful
    if at least 50\% of parties have responded. In this case, the job would contain a configuration
    parameter that triggers aggregation after 10 minutes.

\subsection{End-to-End Illustration}

As illustrated in Figure~\ref{fig:qagg}, a set of parties decide to start an FL job through existing private communication channels.
``Matchmaking'' or inducing parties to join an FL job is out of scope of this paper and \ours.
We assume that this set of parties is convinced of the benfits of FL and want to collaborate.
While forming a group, they also decide things like model architecture, model interchange format
and hyperparameters (initial model weights, 
batch size and learning rate schedule, number of rounds, target accuracy and 
model update frequency). \ours\ then assigns a \textsf{JobID} to this job, creates metadata pertaining
to the job (including party identities and hyperparameters), updates its internal data structures,
instantiates two Kafka queues -- \textsf{JobID-Agg} and \textsf{JobID-Parties}. 
A serverless function is triggered to publish the initial model architecture and weights
on \textsf{JobID-Agg}. The FL job also specifies the triggering function.
Then the first round of training starts at the parties' local
infrastructure using the model downloaded/received from \textsf{JobID-Agg}.

Once local training is complete, parties send model updates to \textsf{JobID-Parties}.
The trigger (serverless) function executes, and if it determines that an aggregation has
to be initiated, triggers a leaf or intermediate aggregator. They pull inputs from \textsf{JobID-Parties}
and publish their outputs to the same. This process continues as model updates arrive. When an aggregator
function determines that all parties have sent their updates, the round is finished and 
the updated model published to \textsf{JobID-Agg}. Then the next round starts.

Job termination criteria may be different depending on the type of the FL job, as discussed earlier. A time-based
or a quorum-based completion criterion may be also used.

\subsection{Durability} Aggregation checkpointing for fault tolerance determines how frequently the aggregator checkpoints its state to external stable storage. 
While this is needed for traditional FL platforms, \ours\ does not use 
    checkpointing. If the execution of a serverless aggregation function fails, it is simply restarted. 
    All aggregator state (updates from parties, partially fused models, etc) is durably stored in message queues.
    This aspect of \ours\ is vital to understanding \ours's resource usage; we observe that the resource overhead of 
    using message queues is equal to that of checkpointing using cloud object stores in single/hierarchical aggregator schemes.

\subsection{Implementation and Elastic Scaling}

We implement \ours\ using the popular Ray~\cite{ray} distributed computing platform. 
Ray provides several abstractions, including powerful serverless functions (Ray remote 
functions). We explored a couple of alternate implementations, including KNative~\cite{knative} and Apache Flink~\cite{flink},
and settled on Ray because it provides arbitrarily long serverless functions, is well integrated 
with common Python libraries (numpy, scikit-learn, Tensorflow and PyTorch) and provides the freedom
to use accelerators if necessary. Ray's internal message queue could have been used
in lieu of Kafka, but we found Kafka to be more robust. Aggregation triggers are implemented
using Ray, and support typical conditions on \textsf{JobID-Parties} (receipt of a certain number of messages, etc.),
but are flexible enough to execute user functions that return booleans (whether aggregation should be triggered or
not).

Our implementation using Ray executes on the Kubernetes cluster manager. Ray's elastic scaler
can request additional Kubernetes pods to execute serverless functions, depending on how
frequently aggregation is triggered. It is also aggressive about releasing unused pods
when there are no model updates pending. When aggregation is triggered, groups of 
model updates are assigned to serverless function invocations. Each invocation is assigned
2 vCPUs and 4GB RAM (this is configurable). If there are insufficient pods to support 
all these invocations, Ray autoscales to request more Kubernetes pods. This also enables
\ours\ to handle large scale party dropouts and joins effectively. Only the exact amount of
compute required for aggregation is deployed -- overheads to spawn tasks on Kubernetes pods 
and create new pods are minimal, as demonstrated in our empirical evaluation.

It is also vital to ensure that model updates are not consumed twice by aggregation functions.
When aggregation is triggered for a model update in a Kafka queue, it as marked using a flag.
The flag is released only after the output of the function is written to Kafka.
If the aggregation function crashes, Ray restarts it, thereby guaranteeing ``exactly once''
processing and aggregation semantics.

\subsection{Expressivity and Security}

The programming model of \ours\ and its implementation using Ray enables us to
support a wide variety of FL aggregation algorithms. Associativity is a pre-requisite for
aggregation scalability; and any associative algorithm can be programmed using
\ours. Most FL aggregation algorithms, including FedAvg/FedSGD~\cite{bonawitz2019towards}, FedProx~\cite{fedprox},
FedMA~\cite{fedma}, Mime~\cite{mime}, Scaffold~\cite{scaffold}, FedPA~\cite{fedpa}, FedPD~\cite{fedpd}
 and  FedDist~\cite{feddist} are associative. In the rare case that the aggregation
algorithm is not associative, \ours\ still uses serverless functions to spawn the single aggregator
instance and does so with a Docker container of the maximum size (configurable) supported by the underlying
Kubernetes cluster. The size and number of aggregator instances, as well as the number of parties handled 
by any single instance are configurable, enabling \ours\ to support FL jobs with varying participation.

Furthermore, none of the design choices of \ours\ has any impact on FL privacy mechanisms used.
Transport layer encryption (TLS) used to transmit model updates in existing FL platforms
can be used to send updates to Kafka in \ours. Updates are decrypted by the aggregation function reading them
from Kafka. \ours\ is oblivious to any noise added by parties
for differential privacy. And the fact that functions in \ours\ can execute most Python code means that
aggregation of homomorphically encrypted model updates (using appropriate libraries) is also feasible.

\section{Evaluation}~\label{sec:eval}

\begin{figure*}[htb]
  \centering
  \includegraphics[width=0.32\textwidth]{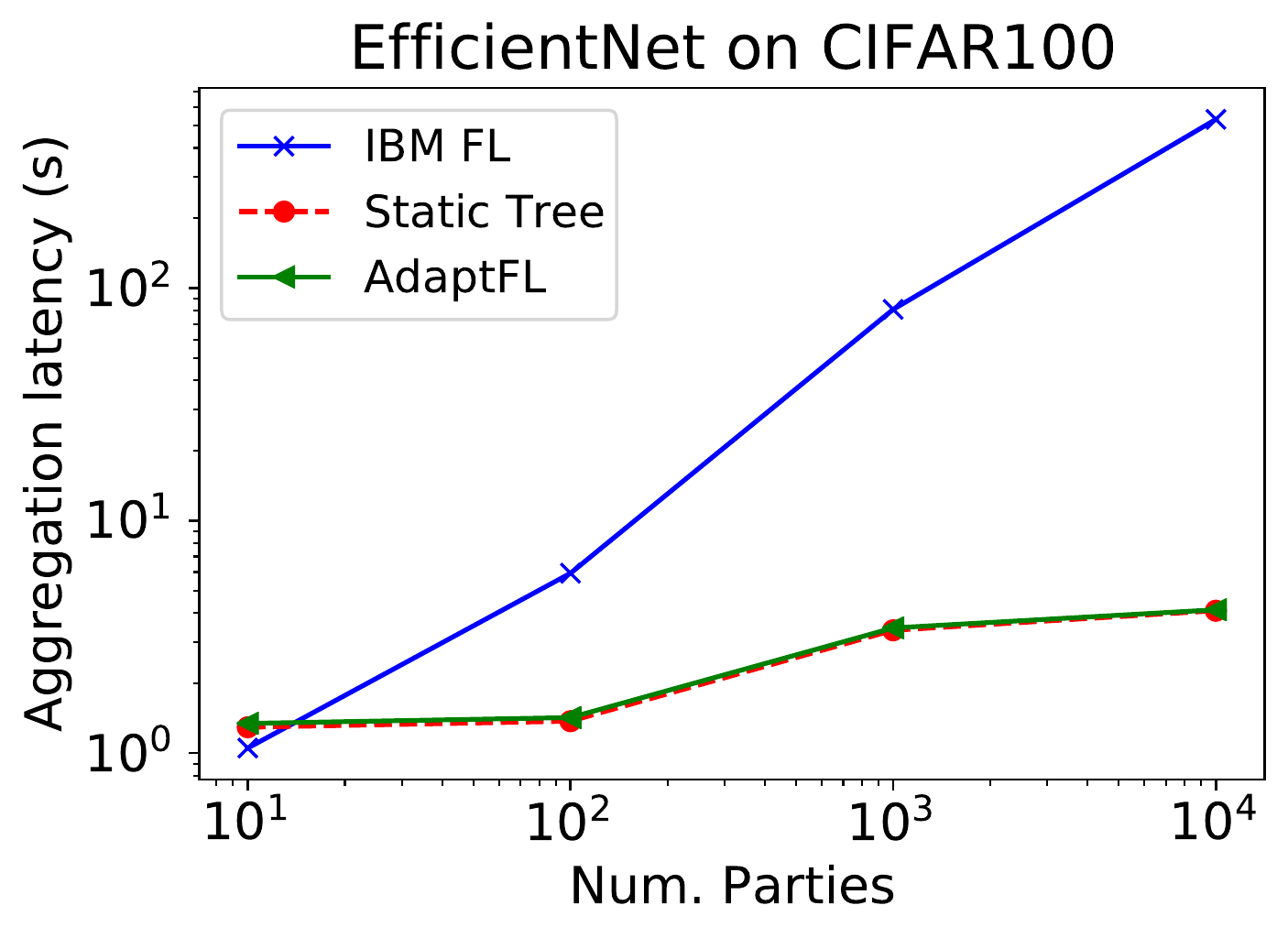}
  \includegraphics[width=0.32\textwidth]{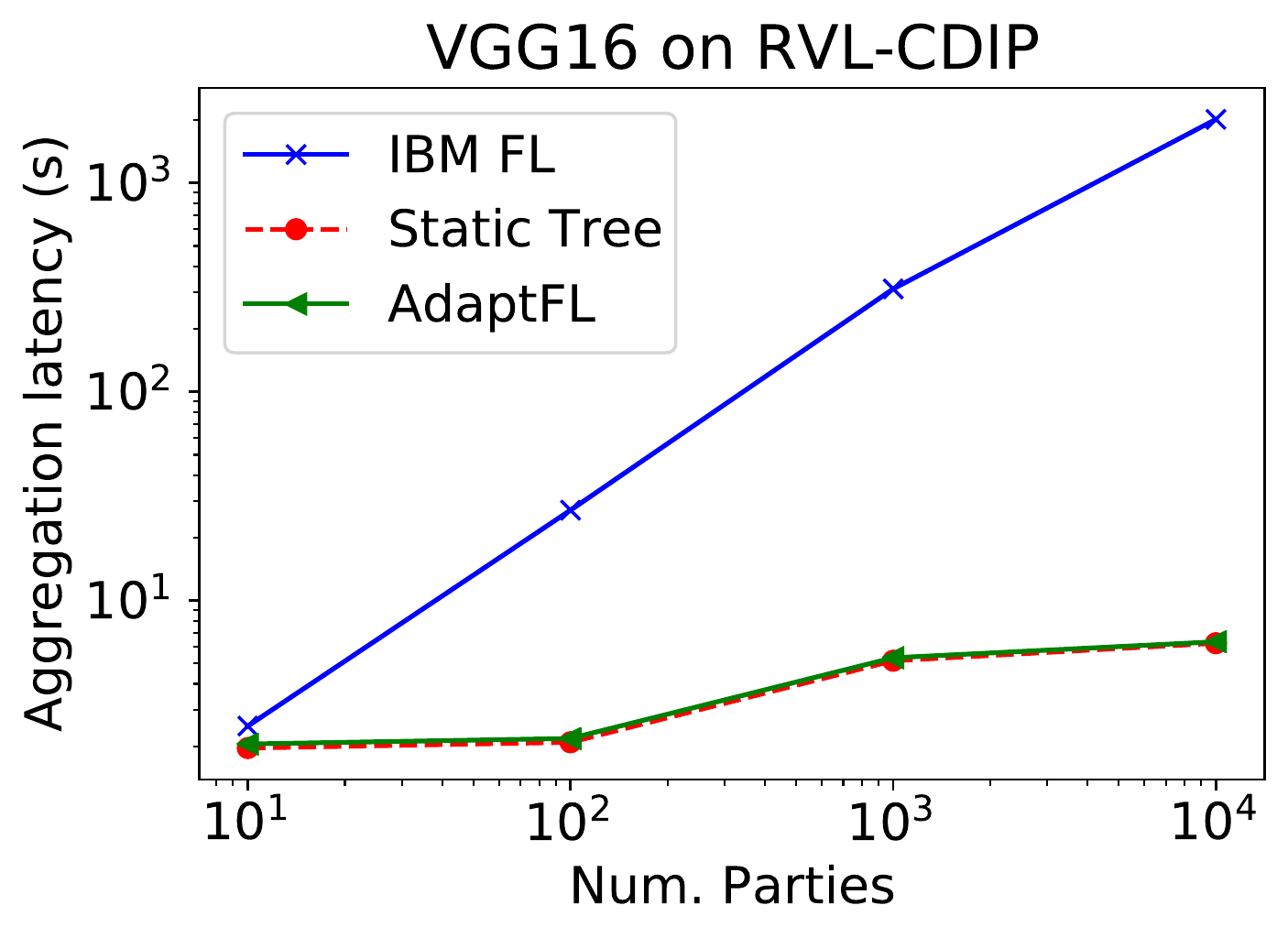}
  \includegraphics[width=0.32\textwidth]{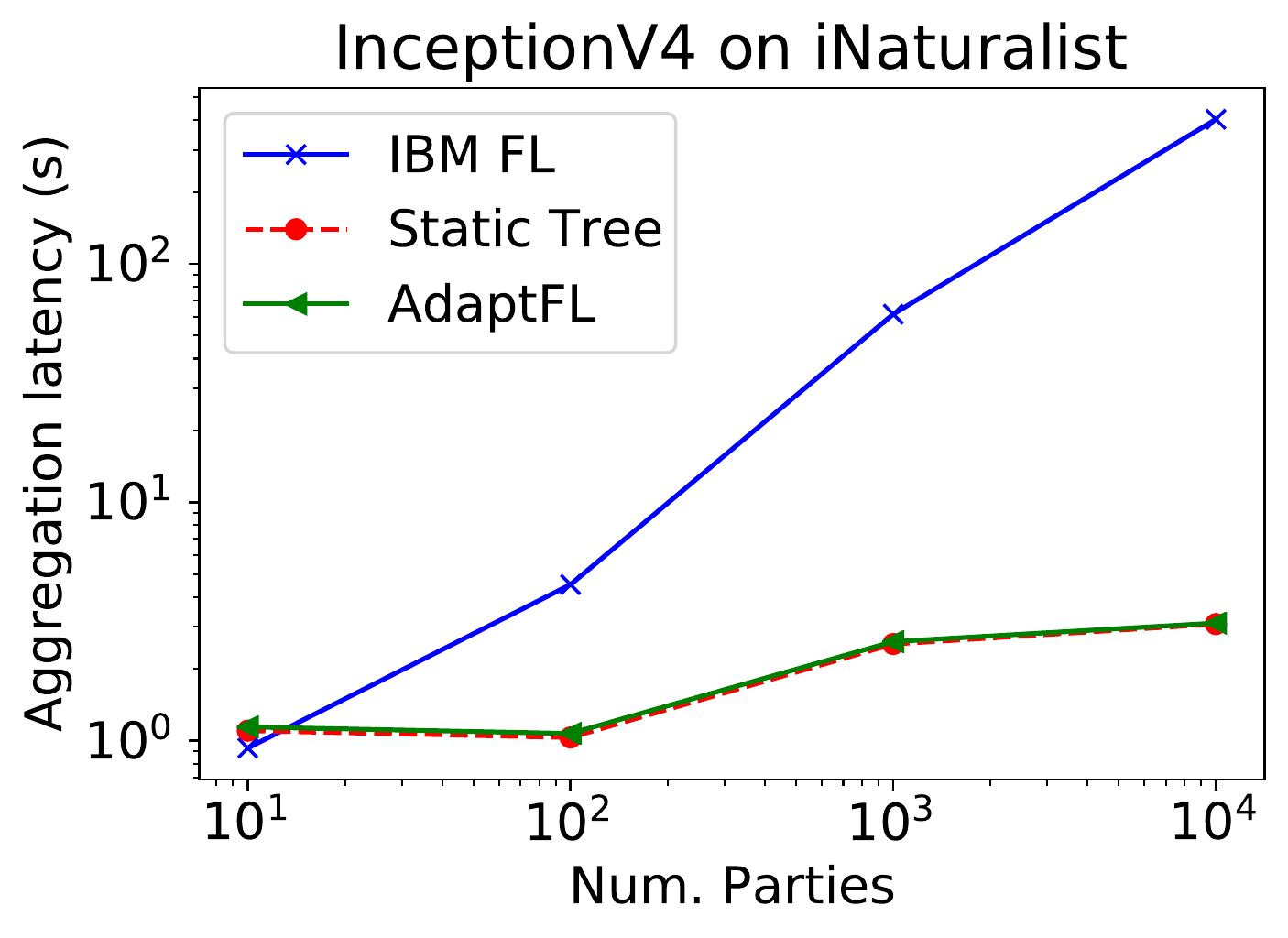}
  \caption{Aggregation Latency (s) -- time taken for aggregation to finish after the last model update is available}
  \label{fig:agglatency-static}
\end{figure*}

\begin{figure}[ht]
  \small
\begin{tabular}{@{}rrrrl@{}}
\toprule
\multicolumn{1}{l}{\# parties} & \multicolumn{1}{l}{Static~Tree (s)} & \multicolumn{1}{l}{Serverless (s)} & \multicolumn{1}{l}{$\frac{Static ~Tree}{Serverless}$} &  \\ \midrule
100   & 4.58  & 1.57 & 2.92$\times$ &  \\
1000  & 12.46 & 4.34 & 2.87$\times$ &  \\
10000 & 15.59 & 4.82 & 3.23$\times$ &  \\ \bottomrule
\end{tabular}
\caption{Effect of 20\% party joins on aggregation latency (seconds). EfficientNet-B7 on CIFAR100 using FedProx aggregation algorithm.}~\label{tbl:joins-cifar}
\end{figure}

\begin{figure}[ht]
  \small
\begin{tabular}{@{}rrrrl@{}}
\toprule
\multicolumn{1}{l}{\# parties} & \multicolumn{1}{l}{Static~Tree (s)} & \multicolumn{1}{l}{Serverless (s)} & \multicolumn{1}{l}{$\frac{Static ~Tree}{Serverless}$} &  \\ \midrule
100   & 10.59 & 4.29 & 2.47$\times$ &  \\
1000  & 17.6  & 6.45 & 2.73$\times$ &  \\
10000 & 26.82 & 7.4  & 3.62$\times$ &  \\ \bottomrule
\end{tabular}
\caption{Effect of 20\% party joins on aggregation latency (seconds). VGG16 on RVL-CDIP using FedSGD aggregation algorithm.}~\label{tbl:joins-rvlcdip}
\end{figure}

\begin{figure}[ht]
\small
\begin{tabular}{@{}rrrrl@{}}
\toprule
\multicolumn{1}{l}{\# parties} & \multicolumn{1}{l}{Static~Tree (s)} & \multicolumn{1}{l}{Serverless (s)} & \multicolumn{1}{l}{$\frac{Static ~Tree}{Serverless}$} &  \\ \midrule
100   & 20.64 & 7.5 & 2.75$\times$ &  \\
1000  & 36.64  & 10.66 & 3.44$\times$ &  \\
7000 & 59.78 & 13.45  & 4.44$\times$ &  \\ \bottomrule
\end{tabular}
\caption{Effect of 20\% party joins on aggregation latency (seconds). InceptionV4 on iNaturalist using FedProx aggregation algorithm.}~\label{tbl:joins-inaturalist}
\end{figure}

In this section, we evaluate the efficacy of \ours, by first comparing \ours\(\)
against a centralized aggregator setup common in several FL
frameworks
like IBM FL~\cite{ibmfl}, FATE~\cite{fate} and NVFLARE~\cite{nvflare}. We demonstrate how such single
aggregator setups have difficulties when scaling beyond 100 participants. We then demonstrate how
a static hierarchical (tree) overlay of aggregator instances can help with the scalability issue,
but is ineffective from a resource consumption, utilization, cost and elasticity perspectives.

\subsection{Metrics}

Given that aggregation depends on \emph{whether} the expected number of model updates are available, we
define \emph{aggregation latency} as the time elapsed between the reception of the last model update
and the availability of the aggregated/fused model. When compared to a static tree deployment of aggregator instances, 
serverless functions are dynamically instantiated in response to model updates. Deployment
of serverless functions takes a small amount of time ($<$ 100 milliseconds) and elastic scaling of
a cluster in response to bursty model update can also take 1-2 seconds. Consequently, the overhead
of aggregation in \ours\ will usually manifest in the form of increased \emph{aggregation latency}.
It is measured for each FL synchronization round,
and the reported numbers in the paper are averaged over all the rounds of the FL job. We want aggregation latency to be as
low as possible. Scalability, or the lack thereof, of any FL aggregation architecture, also manifests in the form
of increased aggregation latency when the number of parties rises.
We therefore evaluate 
(i) \emph{efficiency} by examining whether serverless functions increase the latency
of an FL job, as perceived by a participant, (ii) \emph{scalability} by examining the
impact of the number of parties on latency, (iii) \emph{adaptivity/elasticity}, 
by examining the impact of parties joining midway on latency. 

We evaluate  
\emph{resource efficiency}, by measuring resource consumption (in terms 
of the number and duration of containers used for aggregation), resource (CPU and memory)
utilization and projected total cost.
We execute both hierarchical aggregation and \ours\ using 
containers on Kubernetes pods in our datacenter, and measure the number of \emph{container seconds}
used by an FL job from start to finish. Container seconds is calculated by multiplying the number of 
containers used with the time that each container was used/alive. This includes all the resources used by the ancillary services, 
including MongoDB (for metadata), Kafka and Cloud Object Store. Measuring \emph{container seconds} helps us use
publicly available pricing from cloud providers like Microsoft Azure to project the monetary cost
of aggregation, in both cases, and project cost savings. We also report average CPU and memory utilization,
averaged over the entire FL job.

\subsection{Experimental Setup}

Aggregation was executed on a Kubernetes cluster on CPUs, using Docker containers. For IBM FL, 
the container used for the single aggregator was run on a dedicated server with 16 CPU cores (2.2 Ghz, Intel Xeon 4210) 
and 32GB of RAM. Each container for hierarchical or serverless aggregation
was equipped with 2 vCPUs (2.2 Ghz, Intel Xeon 4210) and 4 GB RAM. For hierarchical/tree aggregation, each
instance was encapsulated using the Kubernetes service abstraction. Parties were emulated, and distributed over
four datacenters (different from the aggregation datacenter) to emulate geographic distribution.
Each party was also executed inside Docker containers (2 vCPUs and 4 GB RAM) on Kubernetes, and these containers
had dedicated resources. We actually had parties running training to emulate realistic federated
learning, as opposed to using, e.g., Tensorflow Federated simulator. 

We select three real-world federated learning jobs -- 
two image classification tasks from the Tensorflow Federated (TFF)~\cite{tff-benchmark} benchmark
and one popular document classification task. From TFF~\cite{tff-benchmark}, we select (i) CIFAR100 dataset which can be 
distributed over 10-10000 parties, with classification performed using the EfficientNet-B7 model and the FedProx~\cite{fedprox}
aggregation algorithm and (ii) iNaturalist dataset
which can be distributed over 10-9237 parties, with classification performed using the InceptionV4
model and FedProx~\cite{fedprox} aggregation algorithm. Thus, 
we consider two types of images and two models of varying sizes. We do not consider other workloads
from TFF because they involve less than 1000 parties. For additional diversity, we consider a third workload
using the VGG16~\cite{vgg16-rvlcdip} model and FedSGD~\cite{bonawitz2019towards} aggrgeation algorithm on RVL-CDIP~\cite{rvlcdip} document classification dataset. Each job was executed for 50 synchronization rounds, with model fusion happening after every local epoch.
For all scenarios, the datasets were partitioned in a realistic non-IID manner.

\subsection{Aggregation Latency and Scalability}~\label{sec:agglatency-scalability}


First, we consider a scenario where the number of parties 
remains constant throughout the FL job, for all synchronization rounds, i.e., once the job starts, no
parties join or leave. 
From Figure~\ref{fig:agglatency-static}, we observe that a centralized
single aggregator setting does not scale to a large number of parties, as average 
aggregation latency increases significantly -- almost linearly.  
This is because of both constrained compute/memory capacity at the single aggregator and 
constrained network bandwidth needed to transfer/load model updates for aggregation.
Figure~\ref{fig:agglatency-static} also illustrates that the increase in aggregation
latency is much more gradual for both static tree overlays and \ours\ (which uses
serverless functions), enabling these architectures to scale to larger FL settings.
In fact, for both static tree and \ours, latency increases only by $\approx~4~\times$
when the number of parties increases 1000$\times$. This trend is due to the data parallelism 
inherent in both the static tree and \ours.

From an efficiency standpoint, we observe that the aggregation latency is similar between static tree and
\ours, within 4\% of each other, with aggregation latency of \ours\ being slightly higher 
than that of the static tree overlay. This is because using serverless functions
does not reduce the number of aggregation steps; it merely avoids having to 
keep the aggregators provisioned and alive when they are not needed. We used runtime profiling to 
determine that the slight (up to 4\%) increase in aggregation latency over the static tree is primarily 
due to cold starts when functions are 
started; the other minor factor is the latency due to the aggregation trigger. Thus, we 
observe that the runtime overhead of using and triggering serverless functions is minimal.

\begin{figure*}[htb]
    \small
    \centering
    \setlength{\tabcolsep}{0.5em}
\begin{tabular}{|r|r|r|r|r|r|r|r|r|r|}
\toprule
\multicolumn{1}{|c|}{Num.} & \multicolumn{2}{|c|}{Tot. container seconds}  & \multicolumn{2}{|c|}{Proj. Total cost US\$} & \multicolumn{1}{|c|}{Cost} 
& \multicolumn{2}{|c|}{Avg. CPU Util. (\%) } & \multicolumn{2}{|c|}{Avg. Memory Util. (\%) } \\
\multicolumn{1}{|c|}{Parties} &
  \multicolumn{1}{|c|}{Static Tree} &
  \multicolumn{1}{|c|}{\ours} &
  \multicolumn{1}{|c|}{Static Tree} &
  \multicolumn{1}{|c|}{\ours} &
  \multicolumn{1}{|c|}{Savings \%} & 
 \multicolumn{1}{|c|}{Static Tree} &
  \multicolumn{1}{|c|}{\ours} &
  \multicolumn{1}{|c|}{Static Tree} &
  \multicolumn{1}{|c|}{\ours}  \\
  \midrule
10    & 1723   & 228 & 0.46  & 0.06 & 86.96\% & 12.31\% & 82.95\% & 46.54\% & 73.35\% \\
100   & 2653   & 351 & 0.71  & 0.09 & 87.32\% & 17.09\% & 83.08\% & 20.89\% & 72.89\% \\
1000  & 22340  & 2951 & 6.01  & 0.79 & 86.86\% & 10.99\% & 83.52\% & 17.23\% & 72.87\% \\
10000 & 298900 & 40849 & 80.46 & 11   & 86.33\% & 10.61\% & 84.27\% & 18.66\% & 75.39\% \\ \bottomrule
\end{tabular}
    \caption{EfficientNet-B7 on CIFAR100 using FedProx aggregation algorithm. Active Participants. Resource usage and projected cost, using container cost/s of 0.0002692 US\$ (source Microsoft Azure\cite{azurepricing})}~\label{tbl:cost-cifar-active}
\end{figure*}

\begin{figure*}[htb]
  \small
    \centering
    \setlength{\tabcolsep}{0.5em}
\begin{tabular}{|r|r|r|r|r|r|r|r|r|r|}
\toprule
\multicolumn{1}{|c|}{Num.} & \multicolumn{2}{|c|}{Tot. container seconds}  & \multicolumn{2}{|c|}{Proj. Total cost US\$} & \multicolumn{1}{|c|}{Cost} 
& \multicolumn{2}{|c|}{Avg. CPU Util. (\%) } & \multicolumn{2}{|c|}{Avg. Memory Util. (\%) } \\
\multicolumn{1}{|c|}{Parties} &
  \multicolumn{1}{|c|}{Static Tree} &
  \multicolumn{1}{|c|}{\ours} &
  \multicolumn{1}{|c|}{Static Tree} &
  \multicolumn{1}{|c|}{\ours} &
  \multicolumn{1}{|c|}{Savings \%} & 
 \multicolumn{1}{|c|}{Static Tree} &
  \multicolumn{1}{|c|}{\ours} &
  \multicolumn{1}{|c|}{Static Tree} &
  \multicolumn{1}{|c|}{\ours}  \\
  \midrule
10    & 1953   & 162   & 0.53  & 0.04 & 91.73\% & 13.17\% & 91.98\% & 47.01\% & 84.36\% \\
100   & 3078   & 234   & 0.83  & 0.06 & 92.4\%  & 10.75\% & 90.22\% & 20.27\% & 82.01\% \\
1000  & 25250  & 1992  & 6.8   & 0.54 & 92.11\% & 13.86\% & 92.92\% & 22.9\%  & 85.82\% \\
10000 & 337830 & 30303 & 90.94 & 8.16 & 91.03\% & 12.36\% & 89.25\% & 22.96\% & 82.89\% \\  \bottomrule
\end{tabular}
\caption{VGG16 on RVL-CDIP using FedSGD aggregation algorithm. Active Participants. Resource usage and projected cost, using 
container cost/s of 0.0002692 US \$ (source Microsoft Azure\cite{azurepricing}}~\label{tbl:cost-rvlcdip-active}
\end{figure*}

\begin{figure*}[htb]
  \small
    \centering
    \setlength{\tabcolsep}{0.5em}
\begin{tabular}{|r|r|r|r|r|r|r|r|r|r|}
\toprule
\multicolumn{1}{|c|}{Num.} & \multicolumn{2}{|c|}{Tot. container seconds}  & \multicolumn{2}{|c|}{Proj. Total cost US\$} & \multicolumn{1}{|c|}{Cost} 
& \multicolumn{2}{|c|}{Avg. CPU Util. (\%) } & \multicolumn{2}{|c|}{Avg. Memory Util. (\%) } \\
\multicolumn{1}{|c|}{Parties} &
  \multicolumn{1}{|c|}{Static Tree} &
  \multicolumn{1}{|c|}{\ours} &
  \multicolumn{1}{|c|}{Static Tree} &
  \multicolumn{1}{|c|}{\ours} &
  \multicolumn{1}{|c|}{Savings \%} & 
 \multicolumn{1}{|c|}{Static Tree} &
  \multicolumn{1}{|c|}{\ours} &
  \multicolumn{1}{|c|}{Static Tree} &
  \multicolumn{1}{|c|}{\ours}  \\
  \midrule
10   & 2365   & 389   & 0.64  & 0.1   & 83.55\% & 10.86\% & 91.86\% & 49.73\% & 82.25\% \\
100  & 3354   & 548   & 0.9   & 0.15  & 83.65\% & 14.17\% & 91.18\% & 21.71\% & 83.49\% \\
1000 & 30545  & 5144  & 8.22  & 1.38  & 83.16\% & 10.87\% & 91.77\% & 23.12\% & 83.43\% \\
9237 & 420870 & 68307 & 113.3 & 18.39 & 83.77\% & 13.44\% & 91.01\% & 21.33\% & 82.49\% \\ \bottomrule
\end{tabular}
\caption{InceptionV4 on iNaturalist using FedProx aggregation algorithm. Active Participants. Resource usage and projected cost, using 
container cost/s of 0.0002692 US \$ (source Microsoft Azure\cite{azurepricing}}~\label{tbl:cost-inaturalist-active}
\end{figure*}

\begin{figure*}[htb]
  \small
    \centering
    \setlength{\tabcolsep}{0.5em}
\begin{tabular}{|r|r|r|r|r|r|r|r|r|r|}
\toprule
\multicolumn{1}{|c|}{Num.} & \multicolumn{2}{|c|}{Tot. container seconds}  & \multicolumn{2}{|c|}{Proj. Total cost US\$} & \multicolumn{1}{|c|}{Cost} 
& \multicolumn{2}{|c|}{Avg. CPU Util. (\%) } & \multicolumn{2}{|c|}{Avg. Memory Util. (\%) } \\
\multicolumn{1}{|c|}{Parties} &
  \multicolumn{1}{|c|}{Static Tree} &
  \multicolumn{1}{|c|}{\ours} &
  \multicolumn{1}{|c|}{Static Tree} &
  \multicolumn{1}{|c|}{\ours} &
  \multicolumn{1}{|c|}{Savings \%} & 
 \multicolumn{1}{|c|}{Static Tree} &
  \multicolumn{1}{|c|}{\ours} &
  \multicolumn{1}{|c|}{Static Tree} &
  \multicolumn{1}{|c|}{\ours}  \\
  \midrule
10    & 634    & 272   & 0.17  & 0.07 & 99.28\% & 10.58\% & 81.3\%  & 42.67\% & 75.26\% \\
100   & 576    & 385   & 0.16  & 0.1  & 98.89\% & 11.97\% & 79.77\% & 12.17\% & 74.77\% \\
1000  & 10516  & 1113  & 2.83  & 0.3  & 99.82\% & 11.41\% & 81.06\% & 11.05\% & 74.15\% \\
10000 & 105021 & 18741 & 28.27 & 5.05 & 99.7\%  & 10.25\% & 81.09\% & 10.29\% & 74.71\% \\ \bottomrule
\end{tabular}
\caption{EfficientNet-B7 on CIFAR100 using FedProx aggregation algorithm. Intermittent participants updating over a 10 minute interval for every synchronization round. Resource usage and projected cost using Container cost/s of 0.0002693 US \$
(source Microsoft Azure~\cite{azurepricing}).}~\label{tbl:cost-cifar-inter}
\end{figure*}

\begin{figure*}[htb]
    \small
    \centering
    \setlength{\tabcolsep}{0.5em}
\begin{tabular}{|r|r|r|r|r|r|r|r|r|r|}
\toprule
\multicolumn{1}{|c|}{Num.} & \multicolumn{2}{|c|}{Tot. container seconds}  & \multicolumn{2}{|c|}{Proj. Total cost US\$} & \multicolumn{1}{|c|}{Cost} 
& \multicolumn{2}{|c|}{Avg. CPU Util. (\%) } & \multicolumn{2}{|c|}{Avg. Memory Util. (\%) } \\
\multicolumn{1}{|c|}{Parties} &
  \multicolumn{1}{|c|}{Static Tree} &
  \multicolumn{1}{|c|}{\ours} &
  \multicolumn{1}{|c|}{Static Tree} &
  \multicolumn{1}{|c|}{\ours} &
  \multicolumn{1}{|c|}{Savings \%} & 
 \multicolumn{1}{|c|}{Static Tree} &
  \multicolumn{1}{|c|}{\ours} &
  \multicolumn{1}{|c|}{Static Tree} &
  \multicolumn{1}{|c|}{\ours}  \\
  \midrule
10    & 33043   & 258    & 8.9     & 0.07  & 99.21\% & 13.23\% & 87.06\% & 46.98\% & 82.11\% \\
100   & 33037   & 385    & 8.89    & 0.1   & 98.88\% & 14.12\% & 84.2\%  & 10.3\%  & 81.56\% \\
1000  & 510039  & 2975   & 137.3   & 0.8   & 99.42\% & 14.46\% & 85.77\% & 10.69\% & 81.7\%  \\
10000 & 5700030 & 40884  & 1534.45 & 11.01 & 99.28\% & 10.91\% & 84.27\% & 12.08\% & 80.86\% \\ \bottomrule
\end{tabular}
\caption{VGG16 on RVL-CDIP using FedSGD aggregation algorithm. Intermittent participants updating over a 10 minute interval for every synchronization round. Resource usage and projected cost using Container cost/s of 0.0002693 US \$
(source Microsoft Azure~\cite{azurepricing}).}~\label{tbl:cost-rvlcdip-inter}
\end{figure*}

\begin{figure*}[htb]
  \small
    \centering
    \setlength{\tabcolsep}{0.5em}
\begin{tabular}{|r|r|r|r|r|r|r|r|r|r|}
\toprule
\multicolumn{1}{|c|}{Num.} & \multicolumn{2}{|c|}{Tot. container seconds}  & \multicolumn{2}{|c|}{Proj. Total cost US\$} & \multicolumn{1}{|c|}{Cost} 
& \multicolumn{2}{|c|}{Avg. CPU Util. (\%) } & \multicolumn{2}{|c|}{Avg. Memory Util. (\%) } \\
\multicolumn{1}{|c|}{Parties} &
  \multicolumn{1}{|c|}{Static Tree} &
  \multicolumn{1}{|c|}{\ours} &
  \multicolumn{1}{|c|}{Static Tree} &
  \multicolumn{1}{|c|}{\ours} &
  \multicolumn{1}{|c|}{Savings \%} & 
 \multicolumn{1}{|c|}{Static Tree} &
  \multicolumn{1}{|c|}{\ours} &
  \multicolumn{1}{|c|}{Static Tree} &
  \multicolumn{1}{|c|}{\ours}  \\
  \midrule
10   & 34365   & 509    & 9.25    & 0.14  & 98.52\% & 13.49\% & 87.75\% & 51.13\% & 84.17\% \\
100  & 34358   & 588    & 9.25    & 0.16  & 98.29\% & 11.08\% & 87.08\% & 11.88\% & 83.72\% \\
1000 & 734456  & 17700  & 197.72  & 4.76  & 97.59\% & 11.59\% & 89.09\% & 10.1\%  & 87.28\% \\
9237 & 6783036 & 206883 & 1825.99 & 55.69 & 96.95\% & 11.43\% & 88.55\% & 11.19\% & 84.4\%  \\ \bottomrule
\end{tabular}
\caption{InceptionV4 on iNaturalist using FedProx aggregation algorithm. Intermittent participants updating over a 10 minute interval for every synchronization round. Resource usage and projected cost using Container cost/s of 0.0002693 US \$
(source Microsoft Azure~\cite{azurepricing}).}~\label{tbl:cost-inaturalist-inter}
\end{figure*}

\subsection{Adaptivity/Elastic Scaling for Party Joins}

Next, we illustrate how \ours\ can handle parties joining in the middle of the job with minimal impact
on aggregation latency. For this, we consider a single synchronization round, and increase the number of
parties by 20\%. Figures~\ref{tbl:joins-cifar},\ref{tbl:joins-rvlcdip} and \ref{tbl:joins-inaturalist} illustrate the aggregation
latency when 20\% more parties send model updates during the synchronization round. 
For these experiments, we only illustrate static tree based overlays and \ours. This is because
Section~\ref{sec:agglatency-scalability} has already demonstrated that centralized aggregators 
do not scale to handle large numbers of parties; the effect of party joins is similar -- aggregation latency
increases almost linearly w.r.t number of parties joining.
Serverless aggregation in \ours\ needs no overlay reconfiguration, while static tree aggregation needs
to add more aggregator instances and reconfigure the tree. This manifests as a significant increase in aggregation latency
(2.47$\times$ to 4.62$\times$). This is due to the fact that the number of serverless function invocations depends
on the aggregation workload, and partially aggregated updates can be stored in message queues. However,
with a tree overlay, new aggregator nodes have to be instantiated and the topology changed. Thus, although
both static tree and serverless aggregation methods are elastic, using serverless functions provides 
significantly better outcomes.

\subsection{Resource Consumption \& Cost}

We compare \ours\ with static tree aggregation in terms of resource usage. Although the single aggregator 
deployment (e.g., using IBM FL) has much lower resource requirements when compared to \ours, it has significantly higher latency
and does not scale. So, we do not consider it in the experiments in this section. 
We first illustrate the resource consumption of experiments where parties participate actively (as defined in Section~\ref{sec:background}). Figures~\ref{tbl:cost-cifar-active},\ref{tbl:cost-rvlcdip-active} and 
\ref{tbl:cost-inaturalist-active} tabulate the resource usage for the three workloads, in terms of 
container seconds and CPU/memory utilization. This data illustrates the real benefits of
using serverless aggregation, with $>85\%$ resource and cost savings for the EfficientNet-B7/CIFAR100/FedProx job,
$>90\%$ for VGG16/RVL-CDIP/FedSGD and $>80\%$ for InceptionV4/iNaturalist/FedProx. 
These savings are significant and are a direct result of the adaptivity of \ours, by deploying
aggregator functions only when needed. Resource wastage due to static tree can also be observed
from the CPU/memory utilization figures, which are consistently low for static tree because aggregator
instances are idle for long periods. We also observe that, while compute resources needed for aggregation increase
with the number of participants for both static tree and serverless aggregation, the amount of resource and cost
savings remains fairly consistent. We use Microsoft Azure's container pricing for illustrative purposes only;
pricing is similar for other cloud providers. 

We stress that the experiments in Figures~\ref{tbl:cost-cifar-active},\ref{tbl:cost-rvlcdip-active} and 
\ref{tbl:cost-inaturalist-active} are \emph{conservative}; they assume active participation. That is, parties have dedicated resources 
to the FL job, parties do not fail in the middle of training, and training on parties for each round 
starts immediately after a global model is published by the aggregator. In realistic scenarios,
parties (e.g., cell phones or laptops or edge devices) perform many functions other than model training,
have other tasks to do and can only be expected to respond over a period of time (response timeout).
Depending on the deployment scenario, this can be anywhere from several minutes to hours.
Figures~\ref{tbl:cost-cifar-inter},\ref{tbl:cost-rvlcdip-inter} and \ref{tbl:cost-inaturalist-inter} demonstrate that  
resource and cost savings are huge ($>99\%$) when response timeout is set to
\emph{a modest} 10 minutes per aggregation round. Real world FL jobs typically use higher response timeouts and
will thus reap enormous benefits. Thus, our experiments reinforce our
confidence that serverless aggregation can lead to significant resource and cost savings
with minimal overhead.

\section{Related Work}


Parallelzing FL aggregation using a hierarchical topology has been 
explored by ~\cite{bonawitz2019towards}, though the design pattern was introduced by and 
early work on datacenter parallel computing~\cite{grama}. While \cite{bonawitz2019towards}
uses hierarchical aggregation, its programming model is different from \ours. Its primary goal is 
scalability and consequently, it deploys long lived actors instead of serverless functions.
\ours\ aims to make FL aggregation resource efficient, elastic in addition to being scalable;
and use off-the-shelf open source software like  Ray, Kafka and Kubernetes.

Another closely related concurrent work is FedLess~\cite{fedless}, which predominantly uses serverless functions 
for the training side (party side) of FL. FedLess is able to use popular serverless technologies
like AWS Lambda, Azure functions and Openwhisk to enable clients/parties on cloud platforms perform
local training and reports interesting results on using FaaS/serverless instead of IaaS (dedicated VMs and containers)
to implement the party side of FL. It also has the ability to run a single aggregator as a cloud function, but does not
have the ability to parallelize aggregation, and does not seem to scale beyond 200 parties (with 25 parties 
updating per FL round, per \cite{fedless}). Our work in \ours\ has the primary goal of parallelizing and
scaling FL aggregation. Fedless~\cite{fedless} also does not adapt aggregation based on party behavior,
and it is unclear whether parties on the edge (phones/tablets) can train using FedLess.

A number of ML frameworks -- Siren~\cite{siren}, Cirrus~\cite{cirrus} and
the work by LambdaML~\cite{jiang-serverless-ml} use serverless functions
for centralized (not federated) ML and DL training.
Siren~\cite{siren} allows users to train models (ML, DL and RL) in the cloud 
using serverless functions with the goal to reduce programmer burden involved
in using traditional ML frameworks and cluster management technologies for
large scale ML jobs. It also contains optimization algorithms to tune training
performance and reduce training cost using serverless functions. 
Cirrus~\cite{cirrus} goes further, supporting end-to-end centralized ML training workflows
and hyperparameter tuning using serverless functions. 
LambdaML~\cite{jiang-serverless-ml} analyzes
the cost-performance trade-offs between IaaS and serverless
for datacenter/cloud hosted centralized ML training.
LambdaML supports various ML and DL optimization algorithms, and
can execute purely using serverless functions or optimize cost using
a hybrid serverless/IaaS strategy. \ours\ differs from Siren, Cirrus and LambdaML in 
significant ways -- Distributed ML (in Siren, Cirrus and LambdaML) is different from FL. Distributed ML involves 
centralizing data at a data center or cloud service and performing training at a central location.
In contrast, with FL, data never leaves a participant. FL's privacy guarantees are much stronger
and trust requirements much lower than that of distributed ML.

The term ``serverless'' has also been used to refer to peer-to-peer (P2P) federated learning, as
in ~\cite{rw1,rw4, flgossip}. In such systems, aggregation happens over a WAN overlay and not in a datacenter.
The first step involves establishing the overlay network, by following existing technologies like
publish/subscribe overlays, peer discovery, etc~\cite{pubsub, streamoverlays}. The next step involves establishing a spanning
tree over the P2P overlay, routing updates along the spanning tree and aggregating at each node on the tree.
Gossip based learning, \cite{flgossip} does not construct overlays but uses gossip-based broadcast algorithms
to deliver and aggregate model updates in a decentralized manner.  While these techniques are scalable and 
(in the case of gossip algorithms) fault tolerant, they do require either (i) that the model be revealed
to more entities during routing, or (ii) homomorphic encryption~\cite{jayaram-cloud2020} which can be challenging both from a key agreement 
and model size explosion standpoints, or (iii) differential privacy~\cite{abadi-diffpriv} which reduces model accuracy in the 
absence of careful hyperparameter tuning.

\section{Conclusions and Future Work}

In this paper, we have presented \ours, a system for adaptive serverless aggregation in federated learning.
We have described the predominant way of parallelizing aggregation using a tree topology and 
examined its shortcomings. We have demonstrated how serverless/cloud functions can be used to
effectively parallelize and scale aggregation while eliminating resource wastage and 
significantly reducing costs. Our experiments using three different model architectures, datasets
and two FL aggregation algorithms demonstrate that the overhead of using serverless functions for aggregation
is minimal, but resource and cost savings are substantial. We also demonstrate that serverless
aggregation can effectively adapt to handle changes in the number of participants in the FL job.

We are currently working to extend this work in two directions: (i) increasing the dependability
and integrity of aggregation using trusted execution environments (TEEs) and (ii) effectively supporting
multi-cloud environments by using service mesh (like Istio) to find the best aggregator function to
route a model update.




\newlength{\bibitemsep}\setlength{\bibitemsep}{.2\baselineskip plus .05\baselineskip minus .05\baselineskip}
\newlength{\bibparskip}\setlength{\bibparskip}{0pt}
\let\oldthebibliography\thebibliography
\renewcommand\thebibliography[1]{%
  \oldthebibliography{#1}%
  \setlength{\parskip}{\bibitemsep}%
  \setlength{\itemsep}{\bibparskip}%
}


\end{document}